# Die Verarbeitung medizinischer Forschungsdaten ohne Einwilligung – Der Korridor zwischen Anonymisierung und Forschungsausnahme in Österreich[*]

*Saskia Kaltenbrunner, Michael Schmidbauer*

**Saskia Kaltenbrunner**, MRes, ist wissenschaftliche Mitarbeiterin am Institut für Innovation und Digitalisierung im Recht der Universität Wien. Sie befasst sich aus interdisziplinärer Perspektive mit der Regulierung Künstlicher Intelligenz in der Medizin.

Mag. **Michael Schmidbauer**, LL.M. (IT-Law) ist wissenschaftlicher Mitarbeiter am Institut für Innovation und Digitalisierung im Recht der Universität Wien. Er beschäftigt sich in diversen Forschungsprojekten mit Datenschutzrecht und der Regulierung künstlicher Intelligenz.

Moderne, datengetriebene medizinische Forschung erfordert die Verarbeitung sensibler Gesundheitsdaten in großem Ausmaß. Diese sind im System der DSGVO jedoch besonders geschützt, weswegen einer rechtssicheren Verarbeitung in der Praxis regelmäßig datenschutzrechtliche Bedenken entgegenstehen. Diese Bedenken bestehen insbesondere bei Verarbeitung sensibler personenbezogener Daten ohne informierte Einwilligung. Dieser Beitrag analysiert daher Möglichkeiten zur Datenverarbeitung im Bereich der medizinischen Forschung fernab der Einwilligung und beschreibt hierfür das rechtliche Rahmenwerk für Anonymisierung der DSGVO, die nationale, österreichische Umsetzung der Forschungsausnahme und ihr Zusammenspiel.

Modern, data-driven medical research requires the processing of sensitive health data on a large scale. However, this data is subject to special protection under the GDPR, which is why processing regularly raises data protection concerns in practice. These concerns are particularly prevalent when sensitive personal data is processed without informed consent. This article analyses options for data processing in the field of medical research without consent and describes the legal framework for anonymisation under the GDPR, the national Austrian implementation of the research exemption, and their interaction.

---

[*] Christian Doppler Laboratory for Machine Learning Driven Precision Medicine, Department of Biomedical Imaging and Image-guided therapy, Medical University Vienna.

The financial support by the Austrian Federal Ministry for Labour and Economy, the National Foundation for Research, Technology and Development and the Christian Doppler Research Association is gratefully acknowledged.

# Inhalt

# 1. Einleitung und Problemaufriss

Für medizinische Forschungsprojekte ist die Einwilligung betroffener Personen gem Art 9 Abs 2 lit a iVm Art 6 Abs1 lit a DSGVO[1] für die Verarbeitung ihrer sensiblen personenbezogenen Daten sowohl aus ethischer[2] als auch aus (datenschutz-)rechtlicher[3] Sicht der Goldstandard.

Eine Einwilligung muss dabei jedoch den strengen Vorgaben der Freiwilligkeit, Bestimmtheit, Information, Verständlichkeit, leichten Zugänglichkeit und klaren und einfachen Sprache genügen.[4] Zusätzlich macht insbesondere die jederzeit drohende Möglichkeit eines Widerrufs der Einwilligung gem Art 7 Abs 3 DSGVO inklusive Löschverpflichtung gem Art 17 Abs 1 lit b DSGVO diese Verarbeitungsgrundlage weniger attraktiv.[5]

Der Zweckbindungsgrundsatz[6] schreibt vor, dass erhobene Daten nur in einer mit im Vorhinein festgelegten Zwecken kompatiblen Weise verwendet werden dürfen.[7] Erlaubte Verarbeitungen sind nach diesem Grundsatz im Forschungskontext expressis verbis zwar weitreichender als bei einer „normalen" Verarbeitung, gänzlich aufgehoben ist der Zweckbindungsgrundsatz jedoch nicht.[8] Eine Rechtsgrundlage, die für die Primärverarbeitung herangezogen wurde, kann nicht pauschal für eine Sekundärnutzung verwendet werden, somit wird letztendlich auch bei Sekundärnutzung eine eigene Rechtsgrundlage oder eine Ausnahme von diesem Grundsatz benötigt.[9]

Im Forschungskontext wurde zusätzlich der von Rechtswegen verlangte Grad der Informiertheit und somit auch der Zweckbindung, auf die Öffnungsklauseln in der DSGVO gestützt, im österreichischen FOG durch die Etablierung von „broad consent" – und zwar als eigene Verarbeitungsgrundlage[10] – verringert.[11] Das ist vor allem für das Training von KI-Modellen im ergebnisoffenen Big-Data Bereich von Bedeutung.[12]

Trotz dieser Überlegungen ist es nicht immer möglich bzw praktikabel, eine Datenverarbeitung im Forschungskontext auf die Einwilligung der betroffenen Personen zu stützen. Gut ersichtlich ist dies bei der Sekundärnutzung von klinisch erhobenen Daten. Primär zur Patientenversorgung erhobene Daten könnten sich ex post als werthaltig für Forschungszwecke erweisen, eine primär gegebene Einwilligung deckt per se jedoch nicht die weitere Forschungstätigkeit ab. Eine nachträgliche Einwilligung für die damit einhergehende

---

[1] Verordnung (EU) 2016/679 des Europäischen Parlaments und des Rates vom 27. April 2016 zum Schutz natürlicher Personen bei der Verarbeitung personenbezogener Daten, zum freien Datenverkehr und zur Aufhebung der Richtlinie 95/46/EG (Datenschutz-Grundverordnung).
[2] Art 25 WMA Deklaration von Helsinki, Ethische Grundsätze für die medizinische Forschung am Menschen (1964 idF 2013). Link: https://www.bundesaerztekammer.de/fileadmin/user_upload/_old-files/downloads/pdf-Ordner/International/Deklaration_von_Helsinki_2013_20190905.pdf.
[3] Zum – wenigstens systematisch zu postulierenden – „Primat der Einwilligung" siehe *Veil*, Die Datenschutz-Grundverordnung: des Kaisers neue Kleider, NVwZ 2018, 686 (688).
[4] DSB 31.07.2018, DSB-D213.642/0002-DSB/2018.
[5] Zu den dahingehenden Problemen beim Einsatz von KI siehe *Cabral*, Forgetful AI: AI and the Right to Erasure under the GDPR, EDPL 2020/3/8.
[6] Art 5 Abs 1 lit b DSGVO.
[7] *Hötzendorfer/Tschohl/Kastelitz* in *Knyrim*, DatKomm Art 5 DSGVO, Rz 20.
[8] *Drepper*, Datenschutzgerechte Wege zur Nutzung von Real World Data, Prävention und Gesundheitsforschung 2022.
[9] *Knotzer*, Wissenschaftliche Forschung und Datenschutz: Eine kritische Analyse ausgewählter Aspekte der österreichischen Rechtslage, ZTR 2018, 202 (210).
[10] *Cepic*, Broad Consent: Die erweiterte Einwilligung in der Forschung, ZD-Aktuell 2021, 05214.
[11] § 2d Abs 3 FOG.
[12] *Busche*, Einführung in die Rechtsfragen der künstlichen Intelligenz, JA 2023, 441 (446).

Verarbeitung stellt sich mitunter in der Praxis als schwer zu erlangen heraus. Daher sollen im Folgenden andere mögliche Verarbeitungsgrundlagen untersucht werden, insbesondere die Verarbeitung nach erfolgter Anonymisierung, mit der versucht wird, dem Anwendungsbereich der DSGVO gänzlich zu entfliehen und die Verarbeitungen aufgrund des Forschungsprivilegs, bei der Datenverarbeitungen aufgrund von gelockerten Verarbeitungsgrundlagen gerechtfertigt werden.[13]

---

[13] So wurden Einwilligung, Forschungsausnahme und Anonymisierung bereits in der Vergangenheit als die drei Hauptverarbeitungsgrundlagen im Bereich Gesundheit und Big Data betrachtet: *Weichert*, Big Data, Gesundheit und der Datenschutz, DuD 2014, 831 (835); *Knyrim*, Big Data: datenschutzrechtliche Lösungsansätze, Dako 2015, 59.

## 2. Anonymisierung
### 2.1. Grundlagen

Die DSGVO als Fundament des europäischen Datenschutzrechts verfügt in ihrem sachlichen Anwendungsbereich über einen binären Zugang: Personenbezogene Daten unterliegen ihrer Regulierung, nicht-personenbezogene Daten unterliegen ihr nicht. Dieser alles-oder-nichts Ansatz macht die Berufung auf das Verarbeiten von nicht-personenbezogenen Daten zu einem lohnenden Unterfangen. Art 4 Z 1 DSGVO definiert personenbezogene Daten als „Informationen, die sich auf eine identifizierte oder identifizierbare natürliche Person beziehen". Vier kumulative Tatbestandsmerkmale können hierbei ermittelt werden: „Informationen", „beziehen", „identifizierte oder identifizierbare" und „natürliche Personen". Der dritte Punkt ist das Kernelement der Anonymisierungsfrage.[14]

Nicht-personenbezogene Daten erfahren in der DSGVO keine Legaldefinition, sie liegen vielmehr vor, wenn eines der oben genannten Tatbestandsmerkmale fehlt. Im Hinblick auf die Frage der Identifizierung bzw Identifizierbarkeit sind zwei Fälle zu unterscheiden: Informationen, welche gänzlich ohne Personenbezug bestehen bzw hergestellt wurden und Informationen, bei denen bestehender Personenbezug nachträglich entfernt wurde. Letzter Prozess wird als Anonymisierung bezeichnet, das Ergebnis dieses Prozesses sind anonymisierte Daten.

Anonymisierte Daten sind strikt von pseudonymisierten Daten abzugrenzen. Bei letzteren handelt es sich gem Art 4 Z 5 DSGVO um Daten, die „[…] ohne Hinzuziehung zusätzlicher Informationen nicht mehr einer spezifischen betroffenen Person zugeordnet werden können, sofern diese zusätzlichen Informationen gesondert aufbewahrt werden und technischen und organisatorischen Maßnahmen unterliegen, die gewährleisten, dass die personenbezogenen Daten nicht einer identifizierten oder identifizierbaren natürlichen Person zugewiesen werden". Bei der Pseudonymisierung sollen also personenbezogene Daten und deren Identitätsmerkmale getrennt werden – technisch, organisatorisch und logisch.[15] Bei der Pseudonymisierung besteht somit immer die Möglichkeit der Identifizierung der betroffenen Personen. Die Daten können als indirekt personenbezogen bezeichnet werden.[16]

In der Praxis bereitet vor allem die Abgrenzung zwischen pseudonymisierten und anonymisierten Daten Schwierigkeiten. Nach den bisher getroffenen Definitionen kann die Möglichkeit einer Re-Identifizierung als entscheidender Unterscheidungspunkt festgehalten werden. Von wem und mit welcher Wahrscheinlichkeit diese Re-Identifizierung durchgeführt werden müsste, um eine Anonymisierung zu erreichen, wird seit Jahren in der rechtswissenschaftlichen Literatur kontrovers diskutiert.[17]

---

[14] *Bygrave/Tosoni* in *Kuner/Bygrave/Docksey*, The EU General Data Protection Regulation (2020), Art 4.
[15] *Burghoff*, Praxisgerechter Umgang mit der Verfremdung personenbezogener Daten, ZD 2023, 658 (659).
[16] *Hödl* in *Knyrim*, DatKomm Art 4 DSGVO, Rz 57.
[17] Meinungsstreitigkeiten zusammenfassend *Eckhardt*, EuGH: Dynamische IP-Adressen und die Grundsatzfrage zum Anwendungsbereich des Datenschutzrechts, ZIIR 2017, 6.

## 2.2. Relative und faktische Anonymisierung

Da Anonymisierung nicht legaldefiniert ist, muss als Ausgangspunkt für die Beurteilung der Abgrenzung zur Pseudonymisierung ErwGr 26 der DSGVO herangezogen werden. Diese nicht unmittelbar rechtsverbindliche Auslegungshilfe beschreibt Identifizierbarkeit und somit Anonymisierung folgendermaßen: „Um festzustellen, ob eine natürliche Person identifizierbar ist, sollten *alle Mittel* berücksichtigt werden, die von dem *Verantwortlichen oder einer anderen Person* nach allgemeinem Ermessen *wahrscheinlich* genutzt werden, um die natürliche Person direkt oder indirekt zu identifizieren, wie beispielsweise das Aussondern. Bei der Feststellung, ob Mittel nach allgemeinem Ermessen wahrscheinlich zur Identifizierung der natürlichen Person genutzt werden, sollten alle *objektiven Faktoren*, wie die Kosten der Identifizierung und der dafür erforderliche Zeitaufwand, herangezogen werden, wobei die zum Zeitpunkt der Verarbeitung verfügbare *Technologie und technologische Entwicklungen* zu berücksichtigen sind".[18]

Auf Basis dieser Definition hat sich über die letzten Jahre ein relativer, risikobasierter Ansatz mit einigen absoluten Elementen in der datenschutzrechtlichen Literatur durchgesetzt, auf den in Folge näher eingegangen werden soll.[19] Vier Ebenen bzw Problemfelder können dahingehend identifiziert werden: Die Personen, auf deren Re-Identifizierungsmöglichkeit abgestellt werden soll, die Legalität der angewandten Methoden, die Wahrscheinlichkeit der Anwendung von möglichen Methoden und Zusatzwissen und das rechtlich geforderte Re-Identifizierungsrisiko.

### 2.2.1. Wer? – Relativer Personenbezug

In Bezug auf die Re-Identifizierung wird die Frage gestellt, welche Personen bzw Institutionen nicht mehr in der Lage sein dürfen, einen Personenbezug herzustellen. Die Rolle, die dabei das Wissen und die Mittel Dritter einnehmen sollen, beschreibt den Unterschied zwischen der absoluten und der relativen Theorie der Anonymisierung.

Vertreter der absoluten Theorie weiten miteinzubeziehendes Wissen und zu berücksichtigende Mittel auf alle denkbaren Stellen aus: Erst, wenn niemand mehr in der Lage wäre, die anonymisierten Daten zu re-identifizieren, soll rechtlich eine Anonymisierung vorliegen, bei der die Daten nicht mehr der DSGVO unterliegen. Dieser Ansatz lässt sich womöglich aus ErwGr 26 DSGVO herauslesen[20] und hat teilweise Anklang in der Literatur gefunden[21], gilt aber in seiner strengen Ausformung heute als überholt.

Das ist im Wesentlichen dem Urteil *EuGH-Breyer*[22] geschuldet, in dem über den Personenbezug von dynamischen IP-Adressen abgesprochen wurde. Dieser wurde nicht abstrakt geprüft, sondern im Hinblick auf jeweilige Verantwortliche, was zur Durchsetzung der

---

[18] Hervorhebung der Verfasser.
[19] Als Einblick in den Meinungsstreit siehe *Conrad/Folkerts*, Anonyme Daten unter der DSGVO - Risiko statt Vorteil?, KuR 2023, 89 (90).
[20] ErwGr 26 spricht von allen Mitteln, welche vom Verantwortlichen oder einer anderen Person genutzt werden.
[21] *Klabunde* in *Ehmann/Sedlmayr*, Datenschutzgrundverordnung² (2020), Art 4, Rz 20.
[22] EuGH 19.10.2016, C-582/14 – Breyer.

Theorie des relativen Personenbezugs beitrug.[23] Mit *EuGH-Scania*, *EuG-SRB/EDS* und neuerdings *EuGH-IAB Europe* liegen drei Folgeentscheidungen auf EU-Ebene vor, die sich mit den Theorien des Personenbezugs in verschiedenen Kontexten befassen. Der Ansatz des relativen Personenbezugs wird somit auf Basis der in *EuGH-Breyer* entwickelten Grundsätze in Bezug auf Fahrzeugidentifikationsnummern[24], Bankdaten[25] und Cookie-Consent-Strings (TC-Strings)[26] bejaht. Dieser Rechtsprechungslinie folgend muss demnach hinsichtlich jedes potentiellen Verantwortlichen geprüft werden, ob ein gegebenes Datum als anonymisiert anzusehen ist oder nicht.

An dieser Stelle ist insbesondere festzuhalten, dass sich durch den – streng gelesenen – relativen Ansatz ergibt, dass ein und dasselbe Datum für eine Stelle personenbezogen sein kann und für die andere nicht. Somit kann durchaus auch konstatiert werden, dass Pseudonymisierung durch Verschlüsselung uU eine anonymisierende Wirkung haben kann – nur nicht gegenüber Personen, die im Besitz des Schlüssels sind.[27] Pseudonymisierung kann somit, dem relativen Ansatz folgend, als „subjektive Anonymisierung" beschrieben werden.[28] Damit gingen auch Erleichterungen für die Stelle, welche pseudonymisierte Daten teilt, einher – die Möglichkeiten der Auftragsverarbeitung bzw gemeinsamen Verantwortlichkeit gem Art 26 und 28 DSGVO samt Pflicht der dementsprechenden vertraglichen Regelung würde entfallen.[29] Jedoch ist auch diese Ansicht nicht unumstritten, vor allem wenn Zusatzwissen in Betracht gezogen wird.[30]

Dieses Ergebnis scheint auf den ersten Blick auch eindeutig zu sein, lässt jedoch die Möglichkeit der Beschaffung von Zusatzinformationen außer Acht. Es ist mit der relativen Theorie zwar festzuhalten, dass Anonymisierung ein subjektives Unterfangen ist, jedoch stimmt es genauso, dass sich einzelne Verantwortliche jederzeit Zusatzwissen über ihre vermeintlich anonymisierten Daten besorgen könnten und daher zurechnen lassen müssten.[31] Wenn dieser Pool an Zusatzwissen und möglichen Re-Identifizierungsmitteln potenziell unendlich groß wäre, müsste im Ergebnis erst recht wieder einem absoluten Ansatz gefolgt werden. Es benötigt daher ein Regulativ, soll der relative Ansatz nicht ad absurdum geführt werden. Daher verschiebt sich die Diskussion um eine Ebene und wirft zwei Fragen auf: Welcher, als erreichbar zu betrachtenden, Mittel und welchen Zusatzwissens darf sich der potentiell Verantwortliche bedienen und wie wahrscheinlich darf deren Einsatz sein?

---

[23] *Conrad/Folkerts*, Anonyme Daten unter der DSGVO - Risiko statt Vorteil?, KuR 2023, 89 (90); auch aus systematischen Gründen, siehe *Burt/Stalla-Bourdillon*, The definition of 'anonymization' is changing in the EU: Here's what that means. Link: https://iapp.org/news/a/the-definition-of-anonymization-is-changing-in-the-eu-heres-what-that-means/.
[24] EuGH 9.11.2023, C-319/22 – Scania, Rz 45.
[25] EuG 26.4.2023, T-557/20 – SRB/EDS, Rz 100.
[26] EuGH 7.3.2024, C-604/22 – IAB Europe, Rz 49.
[27] *Haimberger/Geuer*, Anonymisierende Wirkung der Pseudonymisierung, Dako 2018, 57.
[28] *Ziebarth*, Das Datum als Geisel – Klarnamenspflicht und Nutzeraussperrung bei Facebook, ZD 2013, 375 (377).
[29] EuG 26.4.2023, T-557/20, DSB 2023, 212 (*Seidel*).
[30] *Hofer*, Überlegungen zur anonymisierenden Wirkung der Pseudonymisierung im Außenverhältnis am Beispiel Cloud-Computing, jusIT 2022, 173; *Geuer/Wollmann*, Verarbeitung von pseudonymen Daten mit besonderem Fokus auf Art 26 und 28 DS-GVO, jusIT 2020, 18 (23).
[31] EuGH 7.3.2024, C-604/22 – IAB Europe, Rz 47; EuGH 9.11.2023, C-319/22 – Scania, Rz 48; EuGH 19.10.2016, C-582/14 – Breyer, Rz 45; EuG 26.4.2023, T-557/20 – SRB/EDS, Rz 104.

### 2.2.2. Wie? – Nur legale Methoden

ErwGr 26 DSGVO schweigt zur Wahl der Mittel, welche berücksichtigt werden müssen und scheint eher das Wahrscheinlichkeitserfordernis als entscheidend zu betrachten. Angesichts dieser Tatsache verwundert es, dass in *EuGH-Breyer*, auf die Schlussanträge des Generalanwalts gestützt, ausschließlich auf legale Mittel der Re-Identifizierung abgestellt wird.[32]

Diese Ansicht wird in der Literatur teilweise übermäßig ernst genommen, wenn beispielsweise sogar gefordert wird, ein vertraglich festgelegtes Re-Identifizierungsverbot genügen zu lassen.[33] Diese Ansicht würde in der Praxis zwar großartige Gestaltungsmöglichkeiten eröffnen und eine Anonymisierung durch Verschlüsselung, Datenaustausch und vertraglicher Absicherung leicht ermöglichen. Sie scheint jedoch nicht dem hohen Schutzniveau der DSGVO zu genügen.

Die herrschende Meinung betrachtet das Merkmal der Legalität nicht als alleinigen Ausschlussgrund der Re-Identifizierung. Das Legalitätserfordernis sollte daher nicht als eigene Kategorie diskutiert, sondern im Zuge des Wahrscheinlichkeitserfordernisses berücksichtigt werden.[34] Auch hier gibt es jedoch eine Bandbreite an Meinungen, welche dieses Erfordernis mehr oder weniger ernst nehmen.[35] Die österreichische Rechtsprechung schweigt, soweit ersichtlich, über dieses Problem.

Aus Praxissicht sei daher trotzdem ein vorsichtiger Standpunkt empfohlen, bei dem im Zweifel davon auszugehen ist, dass auch illegale Mittel der Re-Identifizierung berücksichtigt werden müssen.

### 2.2.3. Wahrscheinlichkeit?

ErwGr 26 DSGVO sieht eine Interpretation vor, bei der in Betracht kommende Mittel „nach allgemeinem Ermessen" und „wahrscheinlich" genutzt werden müssen und bringt somit ein flexibles und daher auch unscharfes Element in die Definition anonymisierter Daten. Dieses Element wird für gewöhnlich als der entscheidende Punkt der Anonymisierungsdefinition betrachtet, so auch von der *Art 29-Datenschutzgruppe*, welche darauf hinweist, dass sich die Verantwortlichen bei der Beurteilung einer erfolgreichen Anonymisierung „auf die konkreten Mittel konzentrieren sollten, die für eine Umkehrung der Anonymisierungstechnik erforderlich wären".[36]

*EuGH-Breyer* liefert zu diesen Begriffen, auf Basis des Vorläufers der DSGVO, der Datenschutzrichtlinie (DS-RL)[37], die Überlegung, dass zur Verfügung stehende Mittel „vernünftigerweise" eingesetzt werden müssen, was auf den Wortlaut von ErwGr 26 DS-RL zurückgeht. Dieses Kriterium wird als nicht erfüllt angesehen, wenn eine Identifizierung einen „*unverhältnismäßigen Aufwand* an Zeit, Kosten und Arbeitskräften erfordern würde, so dass das Risiko einer Identifizierung de facto vernachlässigbar erschiene".[38]

---

[32] EuGH 19.10.2016, C-582/14 – Breyer, Rz 46.
[33] *Conrad/Folkerts*, Anonyme Daten unter der DSGVO - Risiko statt Vorteil?, KuR 2023, 89 (91).
[34] *Ziebarth* in *Sydow*, Europäische Datenschutzgrundverordnung, Artikel 4, Rz 37; *Mayrhofer/Leitner/Stadlbauer*, Datenschutzrechtliche Aspekte der Nutzung von Krankenhausinformationssystemen für Forschungszwecke, ZTR 2022, 207 (219).
[35] *Hofer*, Veranstaltungsbericht: 9. Grazer Datenschutz-Gespräche - "Pseudonymisierung und Anonymisierung im Datenschutzrecht", jusIT 2022, 42.
[36] Art 29-Datenschutzgruppe, Stellungnahme 5/2014 zu Anonymisierungstechniken, WP 216, 10.
[37] Richtlinie 95/46/EG des Europäischen Parlaments und des Rates vom 24. Oktober 1995 zum Schutz natürlicher Personen bei der Verarbeitung personenbezogener Daten und zum freien Datenverkehr.

Gewiss verlagert diese Definition das Problem nur um eine Ebene und es stellt sich die berechtigte Frage, was einen unverhältnismäßigen Aufwand darstellen würde. Eine klare Antwort kann weder Literatur noch Rechtsprechung liefern. Bei der Beurteilung der Wahrscheinlichkeit der eingesetzten Mittel wird für gewöhnlich auf den Stand der Technik und erwerbbares Zusatzwissen abgestellt.[39] Diese beiden Komponenten sind unter dem Begriff der „zur Verfügung stehenden Mittel" zu subsumieren. In Bezug auf diese Mittel ist Anonymisierung als ein Vorhaben zu behandeln, welches sorgfältig geplant werden sollte. Dabei sind insbesondere die zur Verwendung stehenden Anonymisierungstechniken und alle potentiell an der Datennutzung beteiligten Stellen in Betracht zu ziehen. Zusätzlich sei noch betont, dass die Absicht der Re-Identifizierung einer Stelle keine Rolle spielt. Es wird vielmehr nur auf deren objektive Möglichkeiten dazu abgestellt.[40] Ein Restrisiko kann durch diesen faktischen Ansatz verbleiben, wobei auch ein gewisser Spielraum für die verarbeitende Stelle möglich ist.[41]

Die an der Datennutzung beteiligten Stellen sind so gering wie nötig zu halten. Einerseits stimmt es, dass nach der relativen Theorie die Anonymisierung für jede Stelle einzeln zu beurteilen ist, andererseits wird durch die Vergrößerung des Kreises der Datennutzer der Erwerb von Zusatzwissen unüberschaubar und somit das Re-Identifizierungsrisiko erhöht.[42] Bei einem Daten austauschenden Konsortium müsste für jede Stelle erwerbbares Zusatzwissen abgeklärt werden – dies laufend und nicht nur einmalig –, was einen enormen Aufwand bedeuten würde. Somit ergibt sich die Überlegung, dass eine ausreichend sichere Beurteilung des Erwerbs von Zusatzwissen nur bewerkstelligen ließe, wenn die anonymisierten Daten nicht geteilt werden würden und bei der Stelle blieben, welche die Anonymisierung technisch durchführt. Insbesondere eine durch eine andere Stelle vorgenommene Anonymisierung auf Auftrag ist in dieser Hinsicht problematisch.[43]

Der Zugang der DSGVO zu den Anonymisierungstechniken gestaltet sich wie der Rest der Verordnung als technisch neutral und offen. Dieser Umstand sorgt erneut für Flexibilität und Rechtsunsicherheit. Ob eine gewählte Anonymisierungstechnik den Anforderungen einer rechtssicheren Anonymisierung genügt, ist ein Problem, welches großes interdisziplinäres Know-How auf rechtlicher und technischer Ebene erfordert.[44] In ihrer Stellungnahme 5/2014 zu Anonymisierungstechniken sieht die *Art 29 Datenschutzgruppe* eine wahrscheinliche Re-Identifizierung bei Herausgreifen, Verknüpfung und Inferenz als gegeben an und betont, dass diese drei Risiken in Betracht zu ziehen sind, um eine „robuste" Anonymisierung zu gewährleisten. Dabei bedeutet Herausgreifen die Möglichkeit, einen identifizierbaren

---

[38] EuGH 19.10.2016, C-582/14 – Breyer, Rz 46; Hervorhebung der Verfasser.
[39] *Roßnagel/Gemini*, Vertrauen in Anonymisierung, ZD 2021, 487 (488).
[40] *Weichert*, Datenschutz im Kontext der medizinischen Nutzung von KI-Systemen: heute und morgen, Zeitschrift für medizinische Ethik 2021, 351 (354).
[41] *Gierschmann*, Gestaltungsmöglichkeiten durch systematisches und risikobasiertes Vorgehen – Was ist schon anonym?, ZD 2021, 482 (485).
[42] *Johannes/Gemini*, Anonymisierung von Patientendaten durch Fremdlabore für Dritte, MedR 2023, 368 (371); *Roßnagel*, Datenlöschung und Anonymisierung, ZD 2021, 188 (191).
[43] *Stiftung Datenschutz*, Praxisleitfaden zum Anonymisieren personenbezogener Daten, 38; Link: https://stiftungdatenschutz.org/fileadmin/Redaktion/Dokumente/Anonymisierung_personenbezogener_Daten /SDS_Studie_Praxisleitfaden-Anonymisieren-Web_01.pdf.
[44] *Gierschmann*, Gestaltungsmöglichkeiten durch systematisches und risikobasiertes Vorgehen – Was ist schon anonym?, ZD 2021, 482 (485).

Datensatz aus einem Datenbestand zu isolieren, Verknüpfbarkeit die Fähigkeit, mindestens zwei Datensätze zusammenzuführen und Inferenz die Möglichkeit, Werte mit einer signifikanten Wahrscheinlichkeit von anderen Werten abzuleiten. Als Lösung werden daraufhin die Anonymisierungskonzepte der Randomisierung und Generalisierung vorgestellt.[45] All diese Überlegungen sind rein systematischer Natur und greifen für die Beurteilung einer gegebenen Anonymisierung zu kurz. Diese müsste durch interdisziplinäre Zusammenarbeit im jeweiligen Kontext genauer geprüft werden.

Eine der größten Unsicherheiten besteht zusätzlich darin, dass eine bereits erfolgte Anonymisierung nicht permanent ausreicht, vielmehr muss durch die offene Formulierung in ErwGr 26 der DSGVO der Stand der Technik ständig verfolgt werden. Damit ist das Abstellen auf Re-Identifizierungstechniken gemeint, zu denen anonymisierte Daten verarbeitende Stellen Zugang haben. Festzuhalten ist daher, dass durch technischen Fortschritt eine bereits erfolgte Anonymisierung rechtlich hinfällig sein kann. Das führt insbesondere bei langfristig aufbewahrten Daten zu großem Compliance-Aufwand, da deren fortdauernde Anonymisierung periodisch überprüft werden müsste.[46]

### 2.2.4 Endpunkt: Zuordnungswahrscheinlichkeit und Re-Identifizierungsrisiko

Hierbei handelt es sich um die finale Überlegung der Anonymisierungsthematik: Wie sicher muss eine Zuordnung nach Berücksichtigung aller einzusetzenden Mittel und allem Zusatzwissen sein, um den rechtlichen Anforderungen zu genügen? Denkmöglich kommen auch hier zwei mögliche Ansätze in Betracht: Es könnte eine eindeutige Zuordnung zu einer Person verlangt werden oder bereits eine gewisse Zuordnungswahrscheinlichkeit ausreichen.

Die österreichische Datenschutzbehörde ist hier überraschend konkret und vertritt, gestützt auf eine Empfehlung der Datenschutzkommission[47] als ihrer Vorgängerin, in einigen Entscheidungen[48] einen klaren Ansatz: Eine Anonymisierung liege bereits vor, wenn eine statistische Gruppe mit gleichen Merkmalen zumindest sechs Personen umfassen würde. Prozentuell ausgedrückt müsste daher nach österreichischer Rechtslage eine Zuordnungswahrscheinlichkeit von 16,67% genügen, um von einer rechtssicheren Anonymisierung sprechen zu können.

---

[45] *Art 29-Datenschutzgruppe*, Stellungnahme 5/2014 zu Anonymisierungstechniken, WP 216, 11; siehe auch *EDSA*, Leitlinien 04/2020 für die Verwendung von Standortdaten und Tools zur Kontaktnachverfolgen im Zusammenhang mit dem Ausbruch von COVID-19, 6.
[46] *Conrad/Folkerts*, Anonyme Daten unter der DSGVO - Risiko statt Vorteil?, KuR 2023, 89 (92).
[47] Empfehlung der Datenschutzkommission vom 22. Mai 2013, GZ K213.180/0021-DSK/2013.
[48] DSB 14.1.2019, DSB-D123.224/0004-DSB/2018; DSB 30.3.2015, DSB-D215.611/0003-DSB/2014.

## 2.3. Kontext der medizinischen Forschung

Oben getroffene Überlegungen sind nicht ohne Weiteres auf alle medizinischen Kontexte anwendbar. So wird oft angemerkt, dass das Entfernen des Personenbezugs auf viele medizinische Datenkategorien gar nicht immer möglich ist.[49] Daraus folgt unter Umständen auch der Ausschluss einer Pseudonymisierung, wenn sich der direkte Personenbezug gar nicht entfernen ließe.

Während die Aggregierung in statistische Daten als vergleichsweise rechtssicher gilt[50], scheinen unstrukturierte Bilddaten kaum einer rechtssicheren Anonymisierung zugänglich zu sein.[51] Eine rechtssichere Anonymisierung genetischer und biometrischer Daten wird von der herrschenden Meinung generell ausgeschlossen.[52]

Oben getroffene Überlegungen spielen vor allem bei dem häufigen Fall eine Rolle, dass ein Krankenanstaltenträger als datenschutzrechtlich Verantwortlicher Gesundheitsdaten durch Verschlüsselung pseudonymisiert und diese ohne den bei ihm bleibenden Pseudonymisierungsschlüssel mit Forschungseinrichtungen teilt. Nach der Theorie des relativen Personenbezugs müsste davon ausgegangen werden, dass diese Daten unter den folgenden Prämissen für die Forschungseinrichtung als anonymisiert gelten sollten: Erstens dürfte diese keine rechtliche Möglichkeit haben, auf den Pseudonymisierungsschlüssel zuzugreifen, wobei dies Merkmal nicht als generelles Ausschlusskriterium zu werten ist, sondern eher im Sinne einer Wahrscheinlichkeitsverringerung. Zweitens dürfte die Forschungseinrichtung über kein Zusatzwissen verfügen, welches ihr erlauben könnte, aus den de-identifizierten Datensätzen einen Personenbezug herzustellen – dieses Kriterium ist ex ante schwer zu beurteilen.[53] Zusätzlich müssen auch verfügbare Technologien der Forschungseinrichtung in Betracht gezogen werden, welche zu einer Re-Identifizierung verwendet werden könnten.

Um Re-Identifizierungsrisiken anonymisierter Daten zu beurteilen, sollen nach *Mayrhofer/Leitner/Stadlbauer* folgende Fragen berücksichtigt werden. Frage 3 muss bei medizinischen Forschungsprojekten grundsätzlich stets bejaht werden[54]:

1. „Welchen objektiven Stellenwert haben die Daten? Und wie hoch ist der wirtschaftliche Wert der Daten? (Objektiver Kosten-Nutzen-Faktor)
2. Mit welcher technischen Methode wurden die Daten anonymisiert und wie hoch ist die Gefahr, dass sich der aktuelle Stand der Technik ändert? Bzw wie sehr wurden die Daten verfälscht/zusammengefasst etc? (Zeitliche Aufwand zum "De-anonymisieren")

---

[49] *Weichert*, Big Data, Gesundheit und der Datenschutz, DuD 2014, 831 (832).
[50] *Sonntag*, Technische Grenzen der Anonymisierung, jusIT 2018, 137 (142).
[51] *Weitzenboeck/Lison/Cyndecka/Langford*, The GDPR and unstructured data: is anonymization possible? IDPL 2022, 184 (199).
[52] *Johannes/Gemmin*, Anonymisierung von Patientendaten durch Fremdlabore für Dritte, MedR 2023, 368 (371).
[53] *Mayrhofer/Leitner/Stadlbauer*, Datenschutzrechtliche Aspekte der Nutzung von Krankenhausinformationssystemen für Forschungszwecke, ZTR 2022, 207 (214); *Haimberger*, Datenschutz in der medizinischen und pharmazeutischen Forschung (2021), 196; *Fuchs*, Anonymisierende Wirkung der Verschlüsselung, Dako 2019, 55.
[54] *Mayrhofer/Leitner/Stadlbauer*, Datenschutzrechtliche Aspekte der Nutzung von Krankenhausinformationssystemen für Forschungszwecke, ZTR 2022, 207 (222).

3. Handelt es sich um sensible Daten nach Art 9 DSGVO?
4. Zu welchem Verwendungszweck werden die Daten anonymisiert? (Forschungsprojekt, Veröffentlichung etc)
5. Gibt es eine breite Anzahl an Personen mit persönlichem Vorwissen (Familienangehörige, Ärzte etc) zu oder im Zusammenhang mit den später verarbeiteten/anonymisierten Daten?
6. Welche anderen Daten gibt es, die verknüpft werden können?
7. Gibt es wissentliche illegale Mittel der sich ein "Angreifer" theoretisch bedienen könnte, um einen Personenbezug der Daten herzustellen?"

Die Dokumentation der Anonymisierung sollte nach *Gierschmann* folgende Punkte beinhalten[55]:

1. „Angestrebte Anonymisierung (Schutzziele);
2. Gegenstand der Prüfung (Erläuterung der Prüfkriterien, Festlegung der Nutzung der anonymisierten Daten, z.B. offene Daten ja/nein);
3. Daten (Originaldaten, Zwecke der anonymisierten Daten, Dauer der Nutzung);
4. Beteiligte (Verantwortlicher, andere Personen, Dritte);
5. Mittel (technische Mittel, sonstige Hilfsmittel);
6. Systematische Beschreibung des Set-up;
7. Anonymisierungsverfahren (Anonymisierungstechniken, Anwendung);
8. Ergebnis der Anonymisierung (Merkmale, Datenstruktur etc.);
9. Kontrollmechanismen für die Anonymisierung;
10. Grad der Anonymisierung (statistische Beurteilung, Korrelation, weitere Kennzahlen);
11. Risikobewertung (Robustheit/Identifizierungspotenzial, Risiko des Herausgreifens, der Verknüpfbarkeit, der Inferenz).
12. Festlegung von Verfahren und Kriterien, wann/wie regelmäßig neue Kriterien ermittelt werden;
13. Vorgehen bei Vorliegen neuer Risiken."

Die herrschende Meinung sieht in der Anonymisierung an sich eine rechtfertigungsbedürftige Datenverarbeitung iSv Art 4 Z 2 DSGVO und fordert somit für diesen Vorgang eine Rechtsgrundlage.[56] Bei Anonymisierung von Gesundheitsdaten als sensible Daten muss daher den Anforderungen von Art 6 Abs 1 DSGVO und Art 9 Abs 2 DSGVO Genüge getan werden. Die DSGVO enthält dabei keine explizite Anonymisierungsbefugnis. Ein geeigneter Erlaubnistatbestand ist jedoch insbesondere in Art 6 Abs 1 lit f DSGVO zu sehen, da davon ausgegangen werden kann, dass die berechtigten Interessen des Verantwortlichen, im

---

[55] *Gierschmann*, Gestaltungsmöglichkeiten durch systematisches und risikobasiertes Vorgehen – Was ist schon anonym?, ZD 2021, 482 (486).
[56] Anderer Ansicht *Thüsing/Rombey*, Anonymisierung an sich ist keine rechtfertigungsbedürftige Datenverarbeitung, ZD 2021, 548.

Forschungskontext den Personenbezug zu entfernen, regelmäßig die Interessen der Betroffenen überwiegen.[57] Daneben soll nach der DSB außerdem eine Anonymisierung einer Löschung der Daten gleichzusetzen sein, weswegen in Art 6 Abs 1 lit c iVm Art 17 Abs 1 lit a DSGVO eine brauchbare Verarbeitungsgrundlage für diesen Vorgang gesehen werden könnte.[58] Diese Gleichsetzung erfolgte in einer Entscheidung, in der eine rechtliche Verpflichtung zur Löschung nicht sensibler Daten bestand und anstelle der Löschung bloß anonymisiert wurde. Der Schluss, dass aus einer generellen Löschungsbefugnis auch eine generelle Anonymisierungsbefugnis für sensible Daten abzuleiten ist, kann uE jedoch daraus nicht gezogen werden.

Ein korrespondierender Ausnahmetatbestand zu Art 6 Abs 1 lit f DSGVO ist im System von Art 9 Abs 2 DSGVO nicht vorgesehen. Daher würde eine Analyse der Rechtsgrundlagen für die Anonymisierung prima facie zum Ergebnis kommen, dass diese nur bei nicht-sensiblen Daten ohne Einwilligung möglich ist und sensible Daten wie beispielsweise Gesundheitsdaten einer einwilligungsfreien Anonymisierung nicht zugänglich sind. Hier tut sich bei näherer Betrachtung ein Wertungswiderspruch auf, der darin besteht, dass bei diesem Ergebnis die Anonymisierung als ein der Datenminimierung dienliches Schutzinstrument für Betroffene bei sensiblen Daten schwieriger durchzuführen wäre als bei nicht-sensiblen Daten. Von einem das Grundrecht auf Datenschutz wahrendem Rechtsrahmen wäre das gegenteilige Ergebnis zu erwarten. Dieser Wertungswiderspruch wird in der Literatur teilweise durch eine teleologische Reduktion von Art 9 Abs 1 DSGVO aufgelöst, indem dessen Anwendung bei der Anonymisierung von sensiblen Daten generell ausgeschlossen wird. Somit wäre bei einer Anonymisierung nur die oben beschriebene Interessenabwägung gem Art 6 Abs 1 lit f DSGVO durchzuführen.[59]

Es ist jedoch anzumerken, dass auch die pauschale Annahme, Betroffene würden durch Anonymisierung ihrer Daten stets einem geringeren Risiko ausgesetzt sein, zu kurz greift. Beim hier beschriebenen Ausgangssachverhalt, bei denen personenbezogene Daten anonymisiert werden, um mit ihnen weiter Forschung betreiben zu können, besteht gerade durch die Anonymisierung und die damit einhergehende Weiterverarbeitung der Daten ein zusätzliches Risiko für die Betroffenen. Das Durchführen einer teleologischen Reduktion zur Sicherung eines hohen Datenschutzniveaus und der damit einhergehende Verzicht auf die Ausnahmetatbestände von Art 9 Abs 2 DSGVO würde somit keineswegs ihr Ziel erreichen. UE müsste daher kontextbezogen zwischen Anonymisierung, welche durch ihre Risikominimierung tatsächlich ausschließlich im Sinne der Betroffenen ist und Anonymisierung, welche durch die damit ermöglichte Weiterverarbeitung und gegebenenfalls Weitergabe der Daten an Dritte eigentlich eine Risikoerhöhung für Betroffene darstellt, unterschieden werden.[60] Daher erscheint ein Übergehen der Schranken des Art 9 Abs 2 DSGVO im gegebenen Kontext der Sekundärnutzung dogmatisch zweifelhaft. Die Identifizierung eines Ausnahmetatbestandes für die Anonymisierung, welcher demnach möglicherweise doch in der Zustimmung der Betroffenen gem Art 9 Abs 2 lit a DSGVO liegen müsste, ist daher aus Gründen der Rechtssicherheit zu empfehlen.

Wird richtigerweise in der Anonymisierung eine Datenverarbeitung erblickt, hat dies die Anwendbarkeit der Betroffenenrechte gem Art 13ff DSGVO zur Folge. Zuvorderst müsste geprüft werden, ob über eine nachträgliche Anonymisierung informiert werden muss. Dies ist nach den in Art 14 Abs 5 lit b DSGVO festgeschriebenen Grundsätzen zu beurteilen. Demnach kann eine üblicherweise verpflichtende Informationserteilung entfallen, wenn diese einen unverhältnismäßigen Aufwand für den Verantwortlichen erfordern würde. Diese Bestimmungen soll expressis verbis insbesondere bei Verarbeitungen zu wissenschaftlichen Forschungszwecken iSd Art 89 Abs 1 DSGVO zur Anwendung kommen. Die Verhältnismäßigkeit ist zwischen dem Aufwand der Informationserteilung auf der einen Seite und den möglichen negativen Folgen ihres Ausbleibens für die Betroffenen auf der anderen Seite zu beurteilen. Durch den Verweis auf geeignete Maßnahmen gem Art 89 Abs 1 DSGVO kann davon ausgegangen werden, dass insbesondere die Anwendung von Anonymisierungstechniken zugunsten der Ausnahme von der Informationserteilung ausschlagen würde.[61] Mit anderen Worten könnte bereits in der Anonymisierung an sich, sofern ordnungsgemäß durchgeführt, ein Vorgang gesehen werden, welcher eine Informationserteilung über sich selbst ausschließen würde.[62] Ein genereller Ausschluss dieser Informationserteilung wäre nach den oben beschriebenen Risiken jedoch überschießend. Insbesondere durch Datenübermittlung und der unsicheren Beständigkeit der Anonymisierung wird eine Gefahr für die Rechte Betroffener nicht generell auszuschließen und demnach die Durchführung dieser Verhältnismäßigkeitsprüfung unumgänglich sein. Auch bei Annahme der Anwendbarkeit dieser Ausnahme müsste gem Art 14 Abs 5 lit b letzter Satz die Information, dass Daten zur Sekundärnutzung anonymisiert werden und welche Garantien hierfür vorgesehen sind, der Öffentlichkeit bereitgestellt werden. Hierfür eignet sich eine passive Information über diesen Umstand auf der öffentlich einsehbaren Website des Verantwortlichen.

---

[57] *Feiler/Forgó*, EU-DSGVO (2022), Art 6, Rz 41.
[58] DSB 5.12.2018, DSB-D123.270/0009-DSB/2018.
[59] *Hornung/Wagner*, Anonymisierung als datenschutzrelevante Verarbeitung?, ZD 2020, 223 (228); *Stürmer*, Löschen durch Anonymisieren?, ZD 2020, 626 (630).
[60] Diese Risikodivergenz betonend *Johannes/Gemin*, Anonymisierung von Patientendaten durch Fremdlabore für Dritte, MedR 2023, 368 (369).

### 2.4. Zusammenfassung

Da Anonymisierung und Pseudonymisierung geeignete Maßnahme iSd Art 89 DSGVO darstellen, sind diese jedenfalls zu berücksichtigen und vorzunehmen, wenn möglich. Darüber hinaus ist aber bei der Beurteilung des Ergebnisses größte Vorsicht geboten. Im Zweifel sollte ob der unsicheren Rechtslage davon ausgegangen werden, dass pseudonymisierte und somit personenbezogene Daten vorliegen, wenn der direkte Personenbezug überhaupt angemessen entfernt werden kann.[63] Selbst, wenn das Ergebnis anfangs als rechtssichere Anonymisierung angesehen werden könnte, müsste durch den Verweis auf den Stand der Technik und der zwingenden Berücksichtigung von neuen Re-Identifizierungstechniken ein beträchtlicher Verwaltungsaufwand eingegangen werden, welcher die praktische Anwendung dieser Art der Sekundärnutzung beträchtlich erschwert. Weiters ist ein Datenaustausch nach der relativen Theorie des Personenbezugs zwar grundsätzlich leichter möglich, dieser kann jedoch zu einer schwer überschaubaren Menge an leicht zu verschaffendem Zusatzwissen führen, welches einer rechtssicheren Anonymisierung abträglich sein könnte.

---

[61] *Art 29-Datenschutzgruppe*, Leitlinien für Transparenz gemäß der Verordnung 2016/679, WP 260 rev.01, 38.
[62] *Stiftung Datenschutz*, Praxisleitfaden zum Anonymisieren personenbezogener Daten, 60; Link: https://stiftungdatenschutz.org/fileadmin/Redaktion/Dokumente/Anonymisierung_personenbezogener_Daten /SDS_Studie_Praxisleitfaden-Anonymisieren-Web_01.pdf.
[63] *Haimberger*, Datenschutz, 194; *Knotzer*, Wissenschaftliche Forschung und Datenschutz, ZTR 2018, 202 (214); *Weichert*, Datenschutz im Kontext der medizinischen Nutzung von KI-Systemen: heute und morgen, Zeitschrift für medizinische Ethik 2021, 351 (355).

Darüber hinaus muss das gewählte Anonymisierungsverfahren und die fortlaufende Überprüfung seiner Wirksamkeit ausreichend genau dokumentiert werden.[64] Der Standpunkt der hierbei eingenommen werden sollte, ist derjenige der überprüfenden Stelle. Ausreichende Dokumentation könnte Bedenken über die Robustheit der Anonymisierung zerstreuen. Die irische Datenschutzbehörde vertritt einen pragmatischen Ansatz und sieht eben diese Dokumentation als wesentlichen Punkt der Anonymisierung an.[65]

Es kann daher festgehalten werden, dass es sich bei der Anonymisierung im Allgemeinen und bei der Anonymisierung im medizinischen Kontext im Besonderen um ein risikoreiches Unterfangen handelt. Die (datenschutz-)rechtlichen Folgen der fälschlichen Annahme von anonymisierten Daten könnte aufgrund des binären Zugangs der DSGVO nicht größer sein, gleichzeitig ist die Grenze zwischen diesen Datenarten noch immer höchst unsicher. Diese Überlegungen legen den Schluss nahe, dass eine (alleinige) Anonymisierung nicht als Lösung der Ausgangsproblemstellung anzusehen ist.

---

[64] *Gierschmann*, Gestaltungsmöglichkeiten durch systematisches und risikobasiertes Vorgehen – Was ist schon anonym?, ZD 2021, 482

[65] *Irish Data Protection Commissioner*, Guidance on Anonymisation and Pseudonymisation, June 2019, 6; Link: https://www.dataprotection.ie/sites/default/files/uploads/2019-06/190614%20Anonymisation%20and%20Pseudonymisation.pdf.

# 3. Forschungsprivileg

## 3.1. Europäische Rahmenbedingungen: Das Forschungsprivileg in der DSGVO

Wie oben bereits beschrieben, wird für die Verarbeitung personenbezogener Daten eine passende Rechtsgrundlage benötigt. Für Forschungsprojekte können dabei gesonderte Regeln bestehen („Forschungsprivileg" bzw „Forschungsausnahme"), welche insbesondere von Interesse sind, wenn keine Einwilligung der Betroffenen vorliegt oder aus anderen Gründen die Einwilligung als Rechtsgrundlage nicht zweckführend ist. Es wird jedoch nicht in der DSGVO „entschieden" ob und mit welchem Inhalt ein derartiges Forschungsprivileg in den Mitgliedsstaaten eingeführt wird. Stattdessen enthält die DSGVO mit Öffnungsklauseln die Möglichkeit für Mitgliedsstaaten auf nationalstaatlicher Ebene ein Forschungsprivileg einzuführen und dieses konkretisieren. Allerdings müssen dabei alle Rahmenbedingungen, welche die DSGVO vorgibt, eingehalten werden, insbesondere die Anforderung an Verhältnismäßigkeit und angemessene Maßnahmen zum Schutz der Grundrechte der Betroffenen (Art 89 Abs 1 DSGVO).

Verhältnismäßigkeit bedeutet hier, dass die Verarbeitung in einem angemessenen Verhältnis zum Ziel stehen muss. Geeigneten Garantien hingegen sind nicht konkret definiert oder aufgelistet, bei der Verarbeitung von besonderen Datkategorien (also z.B. Gesundheitsdaten) wird aber insbesondere die Datenminimierung in Art 89 Abs 1 als geeignete Maßnahme hervorgehoben. Datenminimierung bedeutet, dass die Verarbeitung der Daten „dem Zweck angemessen und erheblich sowie auf das für die Zwecke der Verarbeitung notwendige Maß beschränkt sein" soll. Dieser Grundsatz beruht auf Art 5 Abs 1 lit c DSGVO, kann jedoch insbesondere im Kontext von Machine-Learning schwer zu definieren und zu erfüllen sein, da Forschungsprojekte, welche mit KI-Modellen arbeiten, darauf beruhen, in großem Ausmaß Daten zu sammeln.

Es bedarf auch einer Definition von wissenschaftlichen- bzw. Forschungszwecken. Hierfür kann auf ErwGr 159 DSGVO gestützt festgehalten werden, dass „wissenschaftliche Forschung" weit auszulegen ist und auch kommerzielle Akteure Forschungszwecke verfolgen können. Medizinische Forschung wird explizit als Beispiel von wissenschaftlicher Forschung iSd DSGVO genannt.

Die Verarbeitung von besonders schutzwürdigen Daten („besondere Datenkategorien") wie Gesundheitsdaten für wissenschaftliche Forschungszwecke wird durch Art 9 Abs 2 lit j DSGVO ermöglicht:

*„die Verarbeitung ist auf der Grundlage des Unionsrechts oder des Rechts eines Mitgliedstaats, das in angemessenem Verhältnis zu dem verfolgten Ziel steht, den Wesensgehalt des Rechts auf Datenschutz wahrt und angemessene und spezifische Maßnahmen zur Wahrung der Grundrechte und Interessen der betroffenen Person vorsieht, für im öffentlichen Interesse liegende Archivzwecke, für wissenschaftliche oder historische Forschungszwecke oder für statistische Zwecke gemäß Artikel 89 Absatz 1 erforderlich."*

Im Zusammenhang damit ist Art 89 Abs 1 DSGVO zu lesen, welcher die geeigneten Garantien für die Rechte und Freiheiten der betroffenen Person, technische und organisatorische Maßnahmen hervorhebt und insbesondere den Grundsatz der Datenminimierung und die Pseudonymisierung betont.

Art 89 Abs 2 DSGVO erlaubt zusätzlich die Einschränkung von Betroffenenrechten bei der Datenverarbeitung zu wissenschaftlichen Forschungszwecken durch nationalstaatliche Bestimmungen oder den Unionsgesetzgeber.

### 3.2. Einführung einer nationalen Rechtsgrundlage

Die Verarbeitung gestützt auf einer solchen Forschungsausnahme setzt jedoch auch voraus, dass für die Verarbeitung zu wissenschaftlichen Forschungszwecken eine nationale Rechtsgrundlage besteht. Sowohl als Rechtsgrundlage als auch zur Einschränkung von Betroffenenrechten in der wissenschaftlichen Forschung (inkl. bei der Verarbeitung von besonderen Kategorien personenbezogener Daten), können also Unterschiede zwischen der Rechtslage in unterschiedlichen Mitgliedsstaaten bestehen. Österreichische Datenverarbeitungen zu wissenschaftlichen Forschungszwecken richten sich demnach an die österreichischen Durchführungsbestimmungen zu den oben genannten Normen.

Der österreichische Gesetzgeber hat das Forschungsprivileg durch zwei Durchführungsbestimmungen eingeführt. Grundsätzlich wurde die Grundlage für die Forschungsausnahme in § 7 Datenschutzgesetz[66] eingeführt. Im Zuge einer Novellierung des Forschungsorganisationsgesetzes[67] im Jahr 2018[68] wurde aber auch hier in § 2d Abs 2 die Grundlage für ein Forschungsprivileg geschaffen. Das Zusammenspiel der beiden Rechtsgrundlagen wird seitdem in der Literatur diskutiert und ist nicht eindeutig geklärt.[69]

So nimmt die wahrscheinlich herrschende Meinung in der Literatur an, dass sich aus den ErlRV zu Art 7 Z 7 (§ 2d) ergibt, dass § 2d Abs 2 FOG eine „speziellere Norm" (lex specialis) zum DSG ist. Dies würde bedeuten, dass §2d Abs 2 FOG dem allgemeineren DSG vorgeht und alleinig anzuwenden ist.[70] Es existiert jedoch auch die gegenteilige Meinung, dass es möglich ist, eine Verarbeitung auf beide Rechtsgrundlagen zu stützen.[71] Entgegen der herrschenden Meinung wendet die Datenschutzbehörde weiterhin § 7 DSG, genau genommen § 7 Abs 3 DSG, welcher sich mit Genehmigungen für Datenverarbeitungen beschäftigt, auf wissenschaftliche Forschungszwecke an und geht daher von einer Parallelität von FOG und DSG aus.[72] Es scheint in der Praxis also durchaus möglich, eine Verarbeitung auf jede der beiden Rechtsgrundlagen zu stützen. Die gewählte Rechtsgrundlage für die Datenverarbeitung (DSG oder FOG) hat praktische Auswirkungen, da erforderliche technische und organisatorische Voraussetzungen unterschiedlich sind.

Zusätzlich ist hervorzuheben, dass das DSG von der (in der DSGVO gegebenen) Möglichkeit, die Betroffenenrechte einzuschränken, keinen Gebrauch gemacht, das FOG jedoch schon. Gem § 2d Abs 6 FOG finden die Betroffenenrechte pauschal keine Anwendung, wenn das Forschungsvorhaben dadurch „unmöglich gemacht oder ernsthaft beeinträchtigt wird".

Im Folgenden werden beide Optionen näher erläutert.

---

[66] Bundesgesetz zum Schutz natürlicher Personen bei der Verarbeitung personenbezogener Daten (Datenschutzgesetz – DSG) BGBl. I Nr. 165/1999.
[67] Forschungsorganisationsgesetz BGBl. Nr. 341/1981.
[68] Im Rahmen des Datenschutzanpassungsgesetzes 2018 – Wissenschaft und Forschung (WFDSAG 2018) BGBl I 2018/31.
[69] *Lachmayer/Souhrada-Kirchmayer*, Datenschutzrecht in der wissenschaftlichen Forschung, zfhr 2018, 153 (157); im Ergebnis auch verneinend *Liebenwein/Bittermann*, Datenschutz in Wissenschaft und Forschung und seine österreichische Umsetzung, RdM 2018, 14 (15).
[70] *Bresich/Dopplinger/Dörnhöfer/Kunnert/Riedl*, Datenschutzgesetz (2018) §7 Rz 10; *Haimberger*, Datenschutz, 21; *Feiler/Forgó*, EU-DSGVO, 527; *Jahnel*, Kommentar zur Datenschutz-Grundverordnung, Art 89 DSGVO, Rz 23.
[71] *Gabauer*, Die Verarbeitung personenbezogener Daten zu wissenschaftlichen Forschungszwecken (2019), 119; *Pinggera/Stadler* in *Brameshuber/Blasch*, Jahrbuch Sozialversicherungsrecht (2022), Von verantwortungsvollen Verantwortlichen, 101; *Lachmayer/Souhrada-Kirchmayer*, Datenschutzrecht in der wissenschaftlichen Forschung, zfhr 2018, 153 (157); *Kotschy* in *Jahnel*, Datenschutzrecht (2020), Die Zulässigkeitsvoraussetzungen für Forschungsdatenverarbeitungen nach dem FOG – eine kritische Analyse, 281 (307); *Grimm* in *Aigner/Kletečka/Kletečka-Pulker/Memmer*, Handbuch Medizinrecht Kap. IV.8.
[72] DSB 07.06.2018, DSB-D202.207/0001-DSB/2018.

### 3.3. Datenschutzgesetz

Das Forschungsprivileg beruht hier auf § 7 Abs 1 DSG. Demnach können für „im öffentlichen Interesse liegende Archivzwecke, wissenschaftliche oder historische Forschungszwecke oder statistische Zwecke, die keine personenbezogenen Ergebnisse zum Ziel haben" personenbezogene Daten verarbeitet werden, solange diese:

- öffentlich zugänglich sind (gem § 7 Abs 1 Z 1 DSG), oder

- für andere Untersuchungen oder andere Zwecke zulässigerweise ermittelt wurden (gem § 7 Abs 1 Z 2 DSG), oder

- für den Verantwortlichen pseudonymisierte Daten sind und der Verantwortliche die Identität der Personen nicht mit rechtlich zulässigen Mitteln bestimmen kann (gem § 7 Abs 1 Z 3 DSG).

Im Rahmen des DSG ist der Begriff „wissenschaftliche Forschung" breit zu interpretieren: „Wissenschaftliche Forschung" soll – wie auch schon nach den Erläuterungen zu der Vorgängerbestimmung in § 46 DSG 2000 und wie auch in der DSGVO vorgegeben – nicht einen inhaltlich abgegrenzten Bereich bezeichnen sondern als Bereich verstanden werden, in dem eine bestimmte Methode der Vorgangsweise, nämlich eine „wissenschaftliche", angewendet wird. Wissenschaftliche Forschung ist daher breit zu verstehen – erfasst etwa nicht nur Grundlagenforschung, sondern auch angewandte Forschung – und kann also durch Verantwortliche des öffentlichen oder des privaten Bereichs vorgenommen werden.[73]

Auch Forschungsprojekte, in denen öffentliche und kommerzielle Akteure zusammenarbeiten und die im Bereich der angewandten Forschung angesiedelt sind, fallen demnach grundsätzlich in den Anwendungsbereich von § 7 DSG.

Die dort normierten Erlaubnistatbestände decken drei Szenarien ab, welche im Forschungskontext von Bedeutung sein können. Die Verarbeitung öffentlich zugänglicher Daten nach § 7 Abs 1 Z 1 DSG umfasst unter anderem Daten, welche in öffentlichen Büchern wie dem Grundbuch zugänglich sind oder im Internet veröffentlicht wurden.[74] Die anderen beiden Erlaubnistatbestände sind für die gegenständlichen Zwecke von größerer Bedeutung.

§ 7 Abs 1 Z 2 DSG erlaubt im Wesentlichen die eine Zweckänderung von primär für andere Zwecke erhobenen Daten. Diese Ausnahme vom datenschutzrechtlichen Zweckbindungsgrundsatz steht auch im Einklang mit dem allgemeinen Postulat von Art 5 Abs 1 lit b DSGVO, dass eine Weiterverarbeitung für im öffentlichen Interesse liegende Archivzwecke, für wissenschaftliche oder historische Forschungszwecke oder für statistische Zwecke nicht als unvereinbar mit den ursprünglichen Zwecken gelten soll. Dieser Erlaubnistatbestand würde sich daher für die Verarbeitung im Ausgangsbeispiel eignen, bei dem rein „intern" Forschung betrieben wird und Behandlungsdaten für wissenschaftliche Zwecke verwendet werden sollen.

§ 7 Abs 1 Z 3 DSG erlaubt den oben[75] bereits diskutierten Datenaustausch pseudonymisierter Daten mit Forschungseinrichtungen, bei denen die Verarbeitung „extern" erfolgt und Daten einem Empfänger offengelegt werden, welcher diese zu anderen als den ursprünglichen Zwecken weiterverarbeitet. Geeignete Garantien iSd Art 89 DSGVO sind in diesem Fall die Pseudonymisierung, welche, die relative Anonymisierungstheorie nachbildend, für den Empfänger anonymisierend wirken soll. Als Erleichterung wird hier expressis verbis auf bloß rechtlich zulässige Mittel der Identifizierung Bezug genommen. Nach dieser Bestimmung muss

---

[73] *Bresich/Dopplinger/Dörnhöfer/Kunnert/Riedl*, DSG, 2018.
[74] *Löffler* in *Knyrim*, DatKomm Art 89 DSGVO, Rz 65.
[75] siehe II. 3. Kontext der medizinischen Forschung.

also die Pseudonymisierung zwar keine anonymisierende Wirkung gegenüber Dritten entfalten, führt jedoch zu einer eigenen Verarbeitungsgrundlage. Durch den Rekurs auf rechtlich zulässige Mittel soll es in diesem Fall sogar möglich sein, durch reine vertragliche Vereinbarung eine Re-Identifizierung iSd § 7 Abs 1 Z 3 DSG rechtssicher auszuschließen. *Bresich/Dopplinger/Dörnhöfer/Kunnert/Riedl* empfehlen diese sogar, um die Anforderungen dieser Bestimmung gerecht zu werden.[76] Somit kann, im Gegensatz zur Anonymisierung, unter dieser Bestimmung sehr wohl eine schriftliche Vereinbarung zwischen Übermittler und Empfänger ausreichen, um das Kriterium der rechtlich zulässigen Mittel auszuschließen.

Für den Anwendungsbereich dieser Bestimmung ist festzuhalten, dass sichergestellt werden muss, dass geeignete Garantien iSd Art 89 DSGVO von den Verantwortlichen eingehalten werden. *Lachmayer/Souhrada-Kirschmayer* argumentieren, dass in den Tatbeständen von § 7 Abs 1 DSG eben jene verhältnismäßige Abstufung der Datenverarbeitung und damit zusammenhängenden Maßnahmen schon erfolgt.[77] Bei der Verarbeitung von Daten, welche vom Verantwortlichen bereits für andere Zwecke ermittelt wurden, besteht die Maßnahme darin, dass die Daten die Sphäre des Verantwortlichen nicht verlassen dürfen. Im Fall des § 7 Abs 1 Z 3 DSG, in welchem dies sehr wohl der Fall ist, bestehen die Maßnahmen eben darin, dass die Daten pseudonymisiert sind und der Verantwortliche die Identität der Betroffenen nicht mit rechtlich zulässigen Mitteln bestimmen kann. Zusätzlich gibt es Restriktionen in § 7 Abs 5 DSG, welche besagen, dass der Personenbezug zu verschlüsseln ist, wenn dies mit dem Zweck der Verarbeitung vereinbar ist (falls also ein Forschungsziel auch mit pseudonymisierten Daten zu erreichen ist, so müssen die Daten in pseudonymisierter Form verarbeitet werden) und das allgemein der Personenbezug gänzlich zu beseitigen ist, und somit eine Anonymisierung vorzunehmen ist, sobald er für die wissenschaftliche Arbeit nicht mehr notwendig ist.

Hervorzuheben ist jedoch, dass § 7 Abs 1 DSG nur zur Anwendung kommt, wenn die Ergebnisse einer Datenverarbeitung nicht personenbezogene Ergebnisse zum Ziel haben. Gedeckt ist hier also nicht die Erarbeitung oder gar Veröffentlichung von personenbezogenen Ergebnissen.[78] Verantwortliche, die sich auf diese Rechtsgrundlage stützen wollen, haben also zu evaluieren, ob das jeweilige Forschungsvorhaben auf personenbezogene Ergebnisse zielt. Hier spielen oben[79] getroffene Überlegungen über den Personenbezug von KI-Modellen an sich eine Rolle. Wird dieser bejaht, würde die Anwendung des § 7 Abs 1 DSG gänzlich entfallen. Dieser Personenbezug ist jedoch zu verneinen.

Zusammengefasst müssen für die Anwendung des Forschungsprivilegs nach § 7 Abs 1 DSG folgende Punkte berücksichtigt werden:

1. Die Datenverarbeitung darf **keine personenbezogenen Ergebnisse** zum Ziel haben

2. Verbleiben die Daten beim Primärnutzer, ist eine Sekundärnutzung für andere Zwecke möglich, werden sie mit **Externen** zu Forschungszwecken geteilt, müssen sie in jedem Fall **pseudonymisiert** werden

3. Darüber hinaus müssen die Voraussetzungen von § 7 Abs 5 DSG erfüllt werden, also **nach Möglichkeit pseudonymisiert und anonymisiert** werden

---

[76] *Bresich/Dopplinger/Dörnhöfer/Kunnert/Riedl*, DSG (2018) § 7 Rz 15.
[77] *Lachmayer/Souhrada-Kirchmayer*, Datenschutzrecht, 153 (156).
[78] *Thiele/Wagner*, Praxiskommentar zum Datenschutzgesetz (2022) § 7 Rz 6.
[79] siehe II. 3. Kontext der medizinischen Forschung.

Diese Überlegungen zeigen sehr gut, dass auch im Kontext des Forschungsprivilegs eine Auseinandersetzung mit dem Themenkomplex der Anonymisierung bzw Pseudonymisierung unumgänglich ist.

Der Vollständigkeit halber sei erwähnt, dass das DSG darüber hinaus in § 7 Abs 2 DSG auch andere Verarbeitungsgrundlagen vorsieht. Diese kommen expressis verbis nur subsidiär zu § 7 Abs 1 DSG zur Anwendung, also wenn die Verarbeitung auf personenbezogene Ergebnisse abzielt oder keiner der in § 7 Abs 1 DSG genannten Erlaubnistatbestände zur Anwendung gelangt. Dafür sind wiederum drei Möglichkeiten für Datenverarbeitungen vorgesehen:

- gemäß besonderer gesetzlicher Vorschriften (§ 7 Abs 2 Z 1 DSG)
- mit Einwilligung der betroffenen Personen (§ 7 Abs 2 Z 2 DSG)
- mit Genehmigung der Datenschutzbehörde (§ 7 Abs 2 Z 3 DSG).

Die ersten beiden Möglichkeiten sind bereits in der DSGVO und Sondergesetzen enthalten und eröffnen daher keine neuen Rechtsgrundlagen. Die Genehmigung der Datenschutzbehörde ist hingegen eine zusätzliche Möglichkeit, welche Verantwortliche in Österreich wahrnehmen können. Hierfür muss ein Antrag gem § 7 Abs 3 DSG and die Datenschutzbehörde gestellt werden, welcher, im Falle von Gesundheitsdaten, das wichtige öffentliche Interesse an der Verarbeitung darlegt, die Unmöglichkeit oder Unverhältnismäßigkeit der Einholung einer Einwilligung erklärt und die fachliche Eignung der Verantwortlichen glaubhaft macht.

Wird § 2d FOG als lex specialis gegenüber § 7 Abs 2 Z 3 DSG interpretiert, dann wird die Option, die Genehmigung der Datenschutzbehörde einzuholen, wohl auch von § 2d Abs 7 FOG überlagert. Dass die Genehmigung nach § 7 Abs 2 Z 3 DSG trotzdem in der Praxis weiterhin zur Anwendung kommt, zeigt eben, dass die Datenschutzbehörde von einer Parallelität von FOG und DSG auszugehen scheint.[80]

---

[80] DSB 07.06.2018, DSB-D202.207/0001-DSB/2018.

### 3.4. Forschungsorganisationsgesetz

Im Zuge einer Novellierung des FOG im Jahr 2018 wurden neue Bestimmungen zur Datenverarbeitung zu Forschungszwecken eingeführt. Datenverarbeitung für wissenschaftliche Forschungszwecke ist in § 2d Abs 2 Z 1 FOG normiert:

*„Für Zwecke dieses Bundesgesetzes dürfen wissenschaftliche Einrichtungen (§ 2b Z 12 FOG), insbesondere auf Grundlage des Art. 9 Abs. 2 Buchstabe g, i und j DSGVO, somit*

> *1. sämtliche personenbezogene Daten jedenfalls verarbeiten, insbesondere im Rahmen von Big Data, personalisierter Medizin, biomedizinischer Forschung, Biobanken und der Übermittlung an andere wissenschaftliche Einrichtungen und Auftragsverarbeiter, wenn*
>> *a) anstelle des Namens, bereichsspezifische Personenkennzeichen oder andere eindeutige Identifikatoren zur Zuordnung herangezogen werden oder*
>> *b) die Verarbeitung in pseudonymisierter Form (Art. 4 Nr. 5 DSGVO) erfolgt oder*
>> *c) Veröffentlichungen*
>>> *aa) nicht oder*
>>> *bb) nur in anonymisierter oder pseudonymisierter Form*
>>>> *erfolgen oder*
>> *d) die Verarbeitung ausschließlich zum Zweck der Anonymisierung oder Pseudonymisierung erfolgt und keine Offenlegung direkt personenbezogener Daten an Dritte (Art. 4 Nr. 10 DSGVO) damit verbunden ist."*

Um den Anwendungsbereich dieser Bestimmung zu eröffnen, muss also eine „wissenschaftliche Einrichtung" vorliegen. Diese werden in § 2b Z 12 FOG sehr breit als „natürliche Personen, Personengemeinschaften sowie juristische Personen, die Zwecke gemäß Art. 89 Abs. 1 DSGVO verfolgen, d.h. insbesondere Tätigkeiten der Forschung und experimentelle Entwicklung (Z 10) vornehmen" definiert. Ob Forschung also im universitären, betrieblichen oder außeruniversitären Bereich stattfindet, spielt keine Rolle.[81]

*Bresich/Dopplinger/Dörnhöfer/Kunnert/Riedl* bezeichnen dies als verfassungsrechtlich problematisch, da in der Abwägung zwischen Forschungsfreiheit (und Erwerbsfreiheit) und dem in § 1 DSG normierten Grundrecht auf Datenschutz die Forschung- und Erwerbsfreiheit über Gebühr priorisiert werde.[82] Es ist demnach unklar, ob § 2d Abs 2 FOG einer verfassungsrechtlichen Kontrolle iSe Verhältnismäßigkeitsprüfung standhalten würde. Kritisiert wird dies vor allem die umfangreiche Erlaubnis zur Verarbeitung sensibler Daten, welche im FOG enthalten ist.[83]

Auch die Europarechtskonformität von § 2d Abs 2 FOG kann auf der Basis von Art 9 Abs 2 lit j DSGVO mit *Knotzer* aus den gleichen Gründen in Zweifel gezogen werden. Die Schaffung einer „gewissermaßen absoluten Erlaubnisnorm"[84] passe nicht in das Konzept der Verhältnismäßigkeit und sei schwer mit der dem Europarecht inhärenten

---

[81] *Haimberger*, Datenschutz, 22. Für die Definition von z.B. Big Data oder biomedizinischer Forschung siehe auch *Haimberger*, Datenschutz, 38 und 50. Diese sind jedoch für die Feststellung der Rechtmäßigkeit der Verarbeitung nicht entscheidend.
[82] *Bresich/Dopplinger/Dörnhöfer/Kunnert/Riedl*, DSG (2018) § 7 Rz 9.
[83] *Beimrohr*, Datenverarbeitung in der Forschung - ein Überblick über das neue Forschungsorganisationsgesetz, N@HZ 2018 SNr, 27 (30).
[84] *Knotzer*, Wissenschaftliche Forschung und Datenschutz: Eine kritische Analyse ausgewählter Aspekte der österreichischen Rechtslage, ZTR 2018, 202 (207).

Wesensgehaltstheorie, welche in concreto nur das Wesen des Datenschutzrechts nicht verletzende Ausnahmen erlaubt, in Einklang zu bringen.[85]

Die Verarbeitungsgrundlagen in § 2d Abs 2 Z 1 FOG decken enorm weite Palette an Datenverarbeitungen ab. Wortlaut und Systematik dieser Norm legen den Schluss nahe, es würde ausreichen, sich auf die weitgehendste Alternative zu berufen, um jegliche Datenverarbeitung zu rechtfertigen, da die Tatbestände alternativ, also nebeneinanderstehend, normiert wurden. Die weitgehendste Alternative ist in § 2d Abs 2 Z 1 lit c sublit bb FOG zu sehen, welcher Datenverarbeitung wissenschaftlicher Einrichtungen erlaubt, solange eine Veröffentlichung der Ergebnisse nur in pseudonymisierter bzw anonymisierter Form erfolgt. Bei strenger Wortinterpretation dieser Bestimmung läuft diese Verarbeitungsgrundlage auf eine nahezu uneingeschränkte Verarbeitungserlaubnis hinaus, welche auf jegliche Verarbeitung im wissenschaftlichen Kontext anzuwenden und nur durch Einschränkungen bei einer anschließenden Veröffentlichung begrenzt ist. Selbst eine Übermittlung an Dritte im Zuge eines Forschungsprojektes wäre somit abgedeckt.[86]

Aus teleologischer Sicht wäre zu erwarten, dass die Bestimmung so gelesen werden soll, dass als Verarbeitungsgrundlage, einschließlich Datenaustausch, eine Pseudonymisierung gem § 2d Abs 2 Z 1 lit b FOG ausreicht und bei einer darüber hinausgehenden Veröffentlichung der Ergebnisse auch sichergestellt werden muss, dass diese ordnungsgemäß pseudonymisiert sind und kein direkter Personenbezug hergestellt werden kann. Ähnlich sind beispielsweise die Rahmenbedingungen der MedUni Wien für Forschung mit Biobanken zu lesen.[87] Allerdings ist nach den Gesetzesmaterialien zum FOG das Ergebnis einer weitreichenden, generellen Verarbeitungsgrundlage vom nationalen Gesetzgeber eindeutig intendiert. Mit § 2 Abs 2 Z 1 FOG soll demnach eine „ausdrückliche und generelle Rechtsgrundlage" geschaffen werden, welche spezielle Big Data Forschungsprojekten zugutekommen soll. Bei diesen sei eine Pseudonymisierung nämlich schwierig zu erreichen.[88]

Daher ist verständlich, wenn iSd Verhältnismäßigkeitsprinzips weitreichendere Garantien gefordert werden, als es bei § 7 DSG der Fall ist. Da die Verarbeitung im Rahmen des FOG nicht inhaltlich eingeschränkt ist (sondern pauschal für alle Forschungszwecke erlaubt wird), soll die Verhältnismäßigkeit der Datenverarbeitung durch „geeignete Garantien" gewährleistet werden (wie in Art 9 Abs 2 lit j und Art 89 Abs 2 DSGVO vorgesehen) und der Schwerpunkt liegt demnach auf der Prüfung dieser Garantien.

Genau hier besteht die Schwierigkeit, mit der Forschende konfrontiert sind, wenn sie sich auf § 2d FOG stützen wollen: Das FOG fordert im Gegenzug für die weitgehende Erlaubnis zur Datenverarbeitung von Forschenden auch äußerst weitreichende „angemessene Maßnahmen" ein. Eine demonstrative und daher nicht abschließende[89] Aufzählung dieser Maßnahmen ist in § 2d Abs 1 FOG enthalten (hier zusammengefasst):

---

[85] Ausführlich *Haimberger*, Datenschutz, 110.
[86] *Kotschy*, Die Zulässigkeitsvoraussetzungen für Forschungsdatenverarbeitungen nach dem FOG – eine kritische Analyse in *Jahnel*, Datenschutzrecht (2020), 300.
[87] Link: https://www.meduniwien.ac.at/web/biobank/allgemeine-information/information-fuer-forscherinnen/gesetzlicher-rahmen/.
[88] ErläutRV 68 BlgNR 26. GP 33.
[89] Laut § 2d Abs 1 FOG sollen „*insbesondere* folgende angemessene Maßnahmen" zur Anwendung kommen.

1. Lückenlose Protokollierung von Zugriffen auf Daten, die automationsunterstützt verarbeitet werden.
2. Geheimhaltungspflicht aller Verantwortlichen, Auftragsverarbeiter, deren Mitarbeitern und anderen Personen in einem arbeitnehmerähnlichen Verhältnis.
3. Automationsunterstützte Verarbeitung personenbezogener Daten auf Grundlage dieses Abschnitts darf ausschließlich für die Zwecke des FOG erfolgen.
4. Es dürfen keine über die Verarbeitung an sich hinausgehende Nachteile für Personen, deren Daten verarbeitet werden, entstehen.
5. Verantwortliche müssen für Verarbeitungen nach Abs 2 zusätzlich:
    a. im Internet öffentlich einsehbar auf die Inanspruchnahme dieser Rechtsgrundlage hinweisen;
    b. bei Ausstattung ihrer Daten mit bereichsspezifischen Personenkennzeichen die Namensangaben sowie andere Personenkennzeichen gemäß Art 87 DSGVO abgesehen von den bereichsspezifischen Personenkennzeichen „Forschung" und bereichsspezifischen Personenkennzeichen in verschlüsselter Form löschen;
    d. die Aufgabenverteilung bei der Datenverarbeitung ausdrücklich festlegen,
    a. die Verarbeitung von Daten an das Vorliegen gültiger Aufträge der anordnungsbefugten Organisationseinheiten und Mitarbeiter binden;
    b. alle Mitarbeiter über Datenschutzvorschriften belehren;
    c. die Zutrittsberechtigung von den Räumlichkeiten, in denen die Datenverarbeitung erfolgt, regeln;
    d. die Zugriffsberechtigung auf Daten und Programme und den Schutz vor Zugriff durch Unbefugte regeln;
    e. Vorkehrungen gegen die unbefugte Inbetriebnahme von Datenverarbeitungsgeräte treffen und die Berechtigung zum Betrieb festlegen;
    f. eine Dokumentation über die nach den lit. d bis i getroffenen Maßnahmen führen, um die Kontrolle und Beweissicherung zu erleichtern.

Bei dem Betreiben von Registerforschung muss gem § 2d Abs 1 Z 5 lit c FOG verpflichtend ein Datenschutzbeauftragter bestellt werden. Zusätzlich regeln § 2d Abs 1 Z 5 lit k-m FOG Anforderungen für einen Antrag bei der Datenschutzbehörde und § 2d Abs 1 Z 6 FOG verbietet die Veröffentlichung von Personenkennzeichen.

*Kotschy* argumentiert, dass § 2d Abs 1 FOG leichter verständlich wäre, wenn im Gesetzestext unterschieden würde zwischen jenen Voraussetzungen welche ohnehin durch die DSGVO bereits für alle Formen der Verarbeitung gelten (insb die Konkretisierung der Datensicherheitsmaßnahmen gem Art 32 DSGVO welche hier enthalten sind) und jenen Maßnahmen, welche speziell für die Verarbeitung zu wissenschaftlichen Forschungszwecken gelten (zB § 2d Abs 1 Z 4 FOG).[90] Es wird kritisiert, dass manche der Maßnahmen, insbesondere die lückenlose Protokollierung und die Hinweisverpflichtung im Internet „überschießend (scheinen), zumindest bei entsprechend sicherer Pseudonymisierung".[91] Wie bei allen

---
[90] *Kotschy*, Zulässigkeitsvoraussetzungen, 296.
[91] *Kotschy*, Zulässigkeitsvoraussetzungen, 297.

Schutzmaßnahmen, welche einen hohen administrativen oder finanziellen Aufwand mit sich bringen, kann angenommen werden, dass es für kleinere Einrichtungen wesentlich schwieriger sein wird, diese zu erfüllen. Vor allem die lückenlose Protokollierung bereitet wohl erheblichen administrativen Aufwand. Verarbeitungen auf diesen Abschnitt zu stützen, kann also in der Praxis für manche Forschungszwecke und wissenschaftliche Einrichtungen schwierig sein. Andererseits haben größere Forschungseinrichtungen, bei denen ein derartiges System vorhanden ist, eine gute Grundlage, um auch ohne Einwilligung Forschungsdaten zu verarbeiten.

Es ist auch in Bezug auf das FOG anzumerken, dass Forschungseinrichtungen nicht umhinkommen, sich mit den Thematiken der Anonymisierung und insbesondere Pseudonymisierung zu beschäftigen, wenn sie das volle Potential des Forschungsprivilegs ausschöpfen wollen.

§ 2d Abs 6 FOG kommt Forschenden zusätzlich entgegen, indem normiert wird, dass die Betroffenenrechte gem Art 12ff DSGVO (Auskunft, Berichtigung, Löschung, Einschränkung der Verarbeitung, Datenübertragbarkeit und Widerspruch) nicht zur Anwendung kommen, insoweit deren Inanspruchnahme die verfolgten Forschungszwecke zumindest ernsthaft beeinträchtigt. Diese Bestimmung soll nach *Lachmayer/Souhrada-Kirchmayer* so ausgelegt werden, dass Betroffene von ihrer Beanspruchung in Kenntnis gesetzt werden müssen. In jedem Fall hat aber die wissenschaftliche Einrichtung die erhebliche Beeinträchtigung der verfolgten Forschungszwecke darzulegen.[92]

### 3.5. Zusammenfassung

Die DSGVO eröffnet Mitgliedsstaaten die Möglichkeit, eine Konkretisierung des Forschungsprivilegs auf nationalstaatlicher Ebene einzuführen. Der österreichische Gesetzgeber hat dies auf zwei Arten wahrgenommen. Zum einen wurde eine Privilegierung für Datenverarbeitung zu wissenschaftlichen Forschungszwecken in § 7 DSG eingeführt. Zum anderen wurde durch eine Novellierung des FOG ebenfalls eine Ausnahme für wissenschaftliche Einrichtungen eingeführt. In der rechtswissenschaftlichen Literatur betrachtet die wohl herrschende Meinung § 2d FOG als besondere Norm gegenüber § 7 DSG, weswegen § 2d FOG daher vorrangig zur Anwendung kommen soll. Dadurch gäbe es kaum Anwendungsfälle für §7 DSG. Es besteht jedoch auch die gegenteilige Meinung in der Literatur und die Datenschutzbehörde scheint von einer Parallelität von DSG und FOG auszugehen.

Verantwortliche, die sich auf das DSG stützen, haben in erster Linie zu beachten, dass § 7 Abs 1 DSG nur anwendbar ist, wenn das Forschungsvorhaben keine personenbezogenen Ergebnisse zum Ziel hat. Darüber hinaus müssen sie sich mit Anonymisierungs- und Pseudonymisierungsmöglichkeiten der verarbeiteten Daten nachweislich auseinandersetzen und diese bei Realisierbarkeit durchführen. Sind diese Voraussetzungen erfüllt, erlaubt § 7 Abs 1 DSG eine Sekundärnutzung von primär zulässigerweise erhobenen Daten durch denselben Verantwortlichen und eine externe Verarbeitung bei erfolgreicher Pseudonymisierung.

Verantwortliche, die sich auf § 2d FOG stützen, haben vor allem die geforderten Maßnahmen in § 2d Abs 1 zu beachten. Die oben zusammengefassten Anforderungen sind sehr streng und können vor allem für kleine wissenschaftlichen Einrichtungen in der Praxis schwer zu erfüllen sein. Auch für größere Einrichtungen bereitet beispielsweise die Pflicht zur Zugriffsdokumentation einiges an Mehraufwand. Sind diese Erfordernisse jedoch einmal erfüllt, erlaubt § 2d Abs 2 Z 1 FOG eine denkbar weitreichende Verarbeitung personenbezogener Daten im Forschungskontext, die nur Einschränkungen bei Veröffentlichung der Ergebnisse vorsieht.

---

[92] *Lachmayer/Souhrada-Kirchmayer*, Datenschutzrecht, 153 (157).

Im Vergleich kann daher festgehalten werden, dass das DSG eine engere Datenverarbeitung, jedoch unter weniger strengen Auflagen erlaubt, während das FOG eine weitreichendere Datenverarbeitung unter höherem administrativem Aufwand vorsieht.

Weitere Unterschiede bestehen in der möglichen Wahrnehmung von Betroffenenrechten, deren pauschale Aussetzung das FOG bei erheblicher Beeinträchtigung der Forschungszwecke erlaubt, die unter dem DSG den Betroffenen dem Grunde nach aber in jedem Fall zustehen und deren Anwendung bei dieser Verarbeitungsgrundlage daher einzeln geprüft werden muss.

Am Ende des Datenlebenszyklus stellt sich zudem noch die Frage der Speicherung von bereits verwendeten Daten. Datenaufbewahrung zählt zur guten wissenschaftlichen Praxis, so wird beispielsweise in den diesbezüglichen Richtlinien der MedUni Wien eine Aufbewahrungsdauer von mindestens zehn Jahren vorgeschrieben[93], was im Vergleich mit den Vorgaben einer Aufbewahrungsdauer von mindestens 30 Jahren im Behandlungskontext des Krankenanstaltenrechts[94] nicht lange erscheint. Je nach Forschungsprojekt wäre darüber hinaus daher eine längere Aufbewahrungsdauer, unter anderem aus Beweisgründen für potentielle Haftungsfragen[95], notwendig. Dies würde jedoch ohne konkrete gesetzliche Grundlage in einem Spannungsverhältnis zum Grundsatz der Speicherbegrenzung gem Art 5 Abs 1 lit e DSGVO stehen. Diese Bestimmung sieht zwar Ausnahmen für wissenschaftliche Forschungszwecke vor, wenn geeignete Maßnahmen wie Pseudonymisierung angewendet werden[96], der österreichische Gesetzgeber hat in § 2d Abs 5 FOG dennoch eine unbeschränkte Aufbewahrungsdauer und somit die komplette Aushebelung des Grundsatzes der Speicherbegrenzung für Zwecke der wissenschaftlichen Forschung iSv Art 89 Abs 1 DSGVO normiert. Dahingehend kann auf der Grundlage des Forschungsprivilegs auch dieser Konflikt gelöst werden.

---

[93] Good Scientific Practice-Richtlinien der MedUni Wien, 18. Link: https://www.meduniwien.ac.at/web/fileadmin/content/forschung/pdf/MedUni_Wien_GSP-Good_Scientific_Practice_de_eng_05102021.pdf.
[94] § 10 Abs 1 Z 3 KAKuG.
[95] So verjährt das Recht, Schadersatzansprüche geltend zu machen gem § 1489 ABGB spätestens nach 30 Jahren.
[96] *Wirth*, Die Pflicht zur Löschung von Forschungsdaten – Urheber- und Datenschutzrecht im Widerspruch zu den Erfordernissen guter wissenschaftlicher Praxis?, ZUM 2020, 585 (590).

# 4. Synthese und Empfehlungen

Wenn zu wissenschaftlichen Forschungszwecken personenbezogene Daten verarbeitet werden sollen, können Forschende eine Reihe an Optionen in Betracht ziehen. Zunächst ist festzuhalten, dass die Verarbeitung personenbezogener Daten ohne eine gültige Rechtsgrundlage nicht rechtmäßig ist (Art 6 Abs 1 DSGVO). Eine solche Rechtsgrundlage ist die Einwilligung (Art 9 Abs 2 lit a iVm Art 6 Abs1 lit a DSGVO), welche aus rechtlicher und ethischer Sicht als bedeutendste Verarbeitungsgrundlage angesehen wird. Eine Einwilligung einzuholen kann jedoch aus zahlreichen Gründen nicht möglich oder zweckführend sein. Zwei weitere Optionen wurden daher näher betrachtet.

Zunächst kann es (nicht nur im Forschungskontext) möglich und lohnend sein, den Personenbezug von Daten zu entfernen, diese also zu anonymisieren. Bei einer Anonymisierung darf, anders als bei der Pseudonymisierung, eine Re-Identifizierung nicht möglich sein. Was dabei aus rechtlicher Sicht diese Anforderung erfüllen soll, wird im Wesentlichen versucht durch ErwGr 26 DSGVO, Kommunikationen von nationalen und EU-Institutionen, Rechtsprechung und einer Fülle an Literatur zu beantworten. Mit einiger Sicherheit kann festgehalten werden, dass sich hierbei ein relativer, risikobasierter Ansatz durchgesetzt hat, der einerseits das Re-Identifizierungsrisiko aus der Perspektive der jeweiligen Verarbeiter bewertet, andererseits eben dieses Risiko nicht gleich null sein muss, sondern einer gewissen Verhältnismäßigkeit zugänglich ist. Welche Methoden und die Beschaffung welches Wissens in diesem Sinne einen unverhältnismäßigen Aufwand darstellen, muss im Einzelfall bewertet werden und erfordert große interdisziplinäre technische und rechtliche Expertise. Es bestehen zwar eine Reihe an Orientierungshilfen für Forschende und einige Fragen, die berücksichtigt werden sollten. Die Grenze zwischen anonymen und pseudonymen Daten bleibt jedoch unpräzise.

Sind Daten anonymisiert, so fällt ihre Verarbeitung außerhalb des Anwendungsbereichs der DSGVO. Daten zu anonymisieren kann daher lohnend für Forschungsprojekte sein, da dies Datenverarbeitung ermöglicht, ohne den Anforderungen der DSGVO zu entsprechen, inklusive der Notwendigkeit einer gültigen Rechtsgrundlage zur Verarbeitung und der Möglichkeit, diese mit anderen Stellen zu teilen. Gerade in der medizinischen Forschung ist die Anonymisierung jedoch oft nicht realisierbar. Bei unstrukturierten Bilddaten, biometrischen oder genetischen Daten ist es oft unmöglich, den Personenbezug rechtssicher komplett zu entfernen. Weiters erfordert auch die Anonymisierung nach herrschender Meinung eine Rechtsgrundlage, welche nicht immer klar zu identifizieren ist.

Selbst wenn eine Anonymisierung gelingt, müsste diese ausreichend dokumentiert werden, um mögliche Beschwerden abzuwehren und – administrativ aufwendiger – periodisch überprüft werden, um Fortentwicklungen im Stand der Technik und dem verfügbaren Zusatzwissen gerecht zu werden. Dies trifft insbesondere auf den Forschungsbereich zu, bei dem verwendete Daten oft über Jahre hinweg aufgrund der guten wissenschaftlichen Praxis aufbewahrt werden müssen. Dabei besteht ständig das Risiko, dass bereits anonymisiert geglaubte Daten plötzlich nur mehr als pseudonymisiert gelten und daher der DSGVO unterliegen. Die Option erscheint daher risikoreich und oft unmöglich.

Eine zweite Möglichkeit für die medizinische Forschung ist daher die Verarbeitung personenbezogener Daten aufgrund einer anderen Rechtsgrundlage als der Einwilligung. Die DSGVO eröffnet die Möglichkeit einer Privilegierung der Forschung, diese muss jedoch auf nationalstaatlicher Ebene konkretisiert werden. Der österreichische Gesetzgeber hat dies auf zwei Arten wahrgenommen: einerseits in § 7 DSG, andererseits in § 2d FOG. Das Zusammenspiel beider Normen ist nicht gänzlich geklärt, die herrschende Meinung sieht in § 2d FOG eine besondere Norm zu § 7 DSG, welche dementsprechend vorrangig zur Anwendung kommen soll. Nichtsdestoweniger scheint sich die österreichische Datenschutzbehörde weiterhin auf beide Rechtsgrundlagen zu stützen. Somit ist also auch die hier im Weiteren vertretene Interpretation möglich, dass beide Normen parallel zur Anwendung kommen können.

Das DSG ermöglicht, solange die Verarbeitung keine personenbezogenen Ergebnisse beziehungsweise gar die Veröffentlichung solcher als Ziel hat, die Verarbeitung durch eine Reihe an Tatbeständen. Hier ist gem § 7 Abs 1 Z 2 DSG zuvorderst eine Sekundärnutzung von selbst zulässig erhobenen Daten durch den Verantwortlichen möglich. § 7 Abs 1 Z 3 DSG erlaubt einen Datenaustausch an Externe, insofern die Daten pseudonymisiert wurden – analog zur Theorie der anonymisierenden Wirkung der Pseudonymisierung. Darüber hinaus ist bei Anwendung von § 7 DSG ohnehin dafür zu sorgen, die Durchführbarkeit einer Pseudonymisierung bzw einer Anonymisierung zu prüfen und bei positivem Ergebnis auch zu realisieren.

Das FOG enthält andererseits eine sehr breite Erlaubnis zur Datenverarbeitung für Forschungszwecken, zB zur Verarbeitung sämtlicher Daten in pseudonymer Form (§ 2d Abs 2 Z 1 lit b) und, in strenger Wortinterpretation, sogar jegliche Verarbeitung, solange Veröffentlichungen nur pseudonymisiert erfolgen. Die Verarbeitung ist also kaum inhaltlich eingeschränkt, das FOG fordert jedoch sehr hohe Maßnahmen zum Schutz der Betroffenen, welche in § 2d Abs 1 (nicht abschließend) aufgelistet sind. Einige dieser Maßnahmen, insbesondere die lückenlose Protokollierung, können für Verantwortliche einen erheblichen administrativen und finanziellen Aufwand darstellen.

Als Conclusio ist zunächst einmal festzuhalten, dass eine Datenverarbeitung ohne Einwilligung der Betroffenen im medizinischen Forschungskontext aus datenschutzrechtlicher Sicht durchaus möglich ist. Forschungseinrichtungen steht hierfür eine breite Palette an Gestaltungsmöglichkeiten offen. Insbesondere bei österreichischen Forschungsprojekten wird durch die, komplizierte und daher nicht immer gelungene, Legistik im Bereich der Forschungsausnahmen eine Wahlmöglichkeit eröffnet, die jedoch einiges an Rechtsunsicherheit schafft.

Von einer reinen Anonymisierungsstrategie ist aus datenschutzrechtlicher Sicht grundsätzlich abzuraten. Das folgt sowohl aus der Rechtsunsicherheit bei der ursprünglichen Anonymisierung als auch aus dem fortwährenden administrativen Aufwand, den es benötigt, um diese aufrecht zu erhalten. Insbesondere bei Big Data Projekten, bei denen die erforderlichen Daten über längere Zeit aufbewahrt werden müssen, scheint diese Strategie alleine nicht zielführend zu sein.

Besteht jedoch trotzdem die Notwendigkeit der Datenverarbeitung kann dies, solange es die gute wissenschaftliche Praxis und ethische Überlegungen erlauben, mitunter durchaus auf das Forschungsprivileg gestützt werden. Hierfür erscheint in Österreich das FOG als sicherste Variante, da dessen Anwendbarkeit gesichert ist, es nahezu uneingeschränkte Datenverarbeitung erlaubt und zusätzlich die Möglichkeit bietet, bei Bedarf und mit guter

Begründung, Betroffenenrecht auszusetzen. Der dafür zu betreibende administrative Aufwand ist jedoch enorm und kann möglicherweise nur von großen Forschungseinrichtungen tatsächlich erbracht werden. Ist ein System nach § 2d Abs 1 FOG bereits vorhanden, scheint eine auf das FOG gestützte Datenverarbeitung als die beste Lösung. Die oben behandelte Diskussion um die Verfassungs- und Europarechtskonformität von § 2d Abs 2 FOG ist aus praktischer Sicht kein Grund, diese Bestimmung unangewendet zu lassen. Es erscheint prima facie nicht als fahrlässig, sich bei der Datenverarbeitung an geltendem, nationalem Recht zu orientieren, dessen gesetzgeberische Wille und Zweck eindeutig erkennbar ist.

Ob das DSG parallel dazu überhaupt anwendbar ist, ist noch immer nicht endgültig geklärt. Sollte es aber zur Anwendung kommen, scheint es für die beiden näher behandelten Anwendungsfälle der Sekundärnutzung selbst erhobener Daten und des Datenaustausches mit Forschungseinrichtungen das Mittel der Wahl zu sein, da es weniger administrativen Aufwand verursacht als eine Verarbeitung, welche auf das FOG gestützt wird. Lediglich das Ergebnis dieser Verarbeitung darf auf keine personenbezogenen Daten abzielen.

Verarbeitung auf der Grundlage des Forschungsprivilegs ist allerdings nicht immer möglich. Ist es für das Projekt notwendig, grenzüberschreitende Datenverarbeitungen durchzuführen, könnte eine Verarbeitung, welche auf ein innerstaatliches Forschungsprivileg gestützt wird, in anderen Verarbeitungsstaaten nicht zulässig sein. In solchen Fällen wäre es unumgänglich, vor einem Datenaustausch eine Verarbeitungsgrundlage auf Basis der DSGVO zu identifizieren.[97] Kann diese nicht gefunden werden, wäre in diesen Fällen doch, wenn möglich, eine Anonymisierung als Lösung denkbar.

Es kann festgehalten werden, dass bei beiden Spielarten des Forschungsprivilegs auch Überlegungen bezüglich geeigneter Strategien zur Pseudonymisierung oder Anonymisierung getroffen werden müssen. Als Ausgangspunkt könnte daher die Anonymisierung angestrebt werden und bei fehlender Realisierbarkeit der Weg über das Forschungsprivileg gewählt werden, bei dem auch für pseudonymisierte Daten Verarbeitungsgrundlagen bestehen.

Demnach sollte aus der Perspektive von Forschenden Anonymisierung auf der einen Seite und der Forschungsausnahme auf der anderen Seite nicht als Weggabelung betrachtet werden, bei der nur eine Möglichkeit in Betracht zu ziehen ist, sondern tatsächlich eher als Korridor, bei dem Rechte schützende Maßnahmen wie Anonymisierung und Pseudonymisierung mit datenschutzrechtlichen Privilegien im Bereich der Forschung einhergehen und genutzt werden können, um das volle Potential dieser Bestimmungen auszuschöpfen.

---

[97] *EDSA,* EDPB Document on response to the request from the European Commission for clarifications on the consistent application of the GDPR, focusing on health research, Rz 15.